%% file: root.tex
\documentclass[conference]{IEEEtran}
\IEEEoverridecommandlockouts

\usepackage{url}

\usepackage{breakurl}
\usepackage[breaklinks]{hyperref}

\usepackage{cite}
\usepackage{amsmath,amssymb,amsfonts}
\usepackage{algorithmic}
\usepackage{graphicx}
\usepackage{textcomp}
\usepackage{xcolor}

\def\BibTeX{{\rm B\kern-.05em{\sc i\kern-.025em b}\kern-.08em
    T\kern-.1667em\lower.7ex\hbox{E}\kern-.125emX}}
\begin{document}

\title{Enhanced well-being assessment as basis for the practical implementation of ethical and rights-based normative principles for AI}

\author{\IEEEauthorblockN{Marek Havrda}
\IEEEauthorblockA{\textit{AI Policy and Social Impact} \\
\textit{GoodAI}\\
Prague, Czech Republic \\
marek.havrda@goodai.com}
\and
\IEEEauthorblockN{Bogdana Rakova}
\IEEEauthorblockA{\textit{Responsible AI} \\
\textit{Accenture}\\
San Francisco, USA \\
bogdana.rakova@accenture.com}
}

\maketitle

\begin{abstract}
Artificial Intelligence (AI) has an increasing impact on all areas of people's livelihoods. A detailed look at existing interdisciplinary and transdisciplinary metrics frameworks could bring new insights and enable practitioners to navigate the challenge of understanding and assessing the impact of Autonomous and Intelligent Systems (A/IS). There has been emerging consensus on fundamental ethical and rights-based AI principles proposed by scholars, governments, civil rights organizations, and technology companies. In order to move from principles to real-world implementation, we adopt a lens motivated by regulatory impact assessments and the well-being movement in public policy. Similar to public policy interventions, outcomes of AI systems implementation may have far-reaching complex impacts. In public policy, indicators are only part of a broader toolbox, as metrics inherently lead to gaming and dissolution of incentives and objectives. Similarly, in the case of A/IS, there's a need for a larger toolbox that allows for the iterative assessment of identified impacts, inclusion of new impacts in the analysis, and identification of emerging trade-offs. In this paper, we propose the practical application of an enhanced well-being impact assessment framework for A/IS that could be employed to address ethical and rights-based normative principles in AI. This process could enable a human-centered algorithmically-supported approach to the understanding of the impacts of AI systems. Finally, we propose a new testing infrastructure which would allow for governments, civil rights organizations, and others, to engage in cooperating with A/IS developers towards implementation of enhanced well-being impact assessments.
\end{abstract}

\section{Introduction}

Rising concerns for the implications of AI-enabled technology has inspired numerous academic publications as well as work by journalists where both researchers and practitioners are taking a critical approach to investigating intended effects, harms, malicious use, and unintended consequences of algorithmic systems. Various scholars have raised difficult questions related to the need for rethinking the publication process in machine learning research \cite{bengio, hecht2018s, 10.1145/3290607.3310421} due to potential negative impacts of technology including behavioral addiction, disinformation and online manipulation by micro-targeting based on individual psychology, erosion of democracy, or technological unemployment. Aligned with these efforts, one of the leading AI research conferences has recently introduced a requirement that all paper submissions include a statement of the "potential broader impact of their work, including its ethical aspects and future societal consequences" \cite{neurips_template}. In each research proposal, authors are asked to discuss both positive and negative outcomes of the application of their work including "a) who may benefit from this research, b) who may be put at disadvantage from this research, c) what are the consequences of failure of the system, and d) whether the task/method leverages biases in the data" \cite{neurips_template}. We further aim to enrich practitioners' perspectives by bringing attention to the need for such ex-ante impact assessments to be driven by a comprehensive evaluation approach informed \textit{inter alia} by the development of impact assessments in the policymaking process.

We start our investigation by defining AI broadly as "autonomous or intelligent software when installed into other software and/or hardware systems that are able to exercise independent reasoning, decision-making, intention forming, and motivating skills according to self-defined principles" \cite{ieee2016ethically}. Furthermore, we will use the terms AI and Autonomous and Intelligent Systems (A/IS) interchangeably. Scholars have discussed an almost overwhelming set of guidelines and principles related to understanding and mitigating negative A/IS implications. For example, Mittelstadt et al. \cite{mittelstadt2019ai} bring forth over sixty sets of ethical guidelines, Zeng et al. \cite{zeng2018linking} provide a taxonomy of 74 sets of principles, while Jobin et al. discuss 84 different sets of principles \cite{jobin2019global}. Fjeld et al. \cite{fjeld2020principled} map the ethical and rights-based approaches to principles for AI by identifying eight common themes set forth by 36 different documents. The AI principles discussed in these documents were written by civil society, private sector, government, inter-governmental, and multistakeholder organizations. The common ground themes they identify can be considered to constitute a \textit{normative core} for ethical and rights-based AI principles -  privacy, accountability, safety and security, transparency and explainability, fairness and non-discrimination, human control of technology, professional responsibility, and promotion of human values \cite{fjeld2020principled}. 

Various scholars have also pointed to the difficulties of putting principles to practice \cite{holstein2019improving} and proposed impact assessments as one approach for their adoption \cite{ainowreport2018}. There exist decades of work on impact assessments in the field of public policy that could be leveraged in the case of A/IS, including but not limited to the regulatory, environmental, human rights, and privacy impact assessments. As a key component of the policy or programming cycle in public management, impact assessments can play two roles: (1) ex-ante impact analysis which is part of the planning activity of the policy cycle and (2) ex-post impact assessment - part of the evaluation and management activity of the policy cycle \cite{reportOECD}. Impact evaluations aim to understand to what extent and how a policy intervention corrects the problem it was intended to address. They are conducted through various methods such as randomized control trials, quasi-experimental methods such as differences-in-differences, matching, and regression discontinuity \cite{martinez2011impact}. Leveraging the tools of policy impact analysis, well-being impact assessments (WIA) could help A/IS developers by providing ex-ante and and ex-post means for the analysis of the A/IS impacts on well-being \cite{schiff2020ieee}. As discussed by Schiff et al. \cite{schiff2020ieee}, the well-being impact assessment involves the iterative implementation of (1) internal analysis, informed by user and stakeholder engagement, (2) development and refinement of a well-being indicator dashboard, (3) data planning and collection, (4) data analysis of the evaluation outputs that could inform improvements for the A/IS.

The main contribution of our work is providing a practical framework for transparent implementation of the \textit{normative core} of AI ethics principles through deployment of well-being impact assessment mechanisms \cite{schiff2020ieee}.
We posit that the impact assessment approach could help in the joint monitoring and assessment of A/IS impacts related to fundamental ethics and rights-based AI principles. Recognizing the challenges of any single actor to implement many of the previously proposed recommendations, we propose an Enhanced Well-Being Impact Assessment (EWIA) to be put in practice through a collaborative multi-stakeholder effort. The EWIA could be executed by establishing joint monitoring and testing systems allowing the collective implementation of AI principles in practice. Well aware of the need to attribute the impact to an A/IS building on social sciences methodologies, we outline an approach to help establish causal links between impacts and an A/IS. Ultimately, we aim to enrich practitioners' toolbox towards ensuring positive A/IS outcomes.

\section{Related Work}

Various scholars have worked on investigating the challenges of governing AI systems \cite{cath2018governing} as well as the development of AI ethics tools, methods, and strategies to translate principles into practices \cite{morley2019initial}. Crawford et al. \cite{ainowreport} provide an overview of the emerging and urgent concerns as well as concrete recommendations for regulators, governments, AI industry, and researchers. Some of the urgent concerns raised by scholars include surveillance, data colonialism, the use of data and algorithms in criminal justice, law enforcement, employment, and healthcare. Various scholars have also proposed checklist approaches \cite{madaio2020co} as well as end-to-end algorithmic auditing frameworks \cite{raji2020closing}. The growing interdisciplinary community of fairness, accountability, and transparency of AI researchers is increasingly investigating the sociotechnical aspects of A/IS through the lens of algorithmic impact assessments. Work by Reisman et al. \cite{ainowreport2018} introduced an algorithmic impact assessment process drawing from impact assessment frameworks in environmental protection, data protection, privacy, and human rights policy domains \cite{ortolano1995environmental,united2012guiding,dpia,bamberger2008privacy}. Selbst et al. situate the algorithmic impact assessment in a sociotechnical context outlining "five 'traps' that [impact assessment] fair-ML work can fall into even as it attempts to be more context-aware in comparison to traditional data science" \cite{selbst2019fairness}. Drawing on studies of sociotechnical systems in Science and Technology Studies, these include the framing trap, portability trap, formalism trap, ripple effect trap, and solutionism trap \cite{selbst2019fairness}, which jointly point to the need for comprehensive evaluation frameworks. Various scholars have proposed impact assessment producing artifacts that aim to guide A/IS creators in the evaluation process. Prior research efforts have primarily been focused on assessment during two stages of the A/IS life-cycle \cite{/content/paper/d62f618a-en} - (1) evaluations of the data on which the system is operating on \cite{gebru2018datasheets,holland2018dataset} and (2) evaluations of the A/IS outcomes \cite{ailabel}. We aim to build on these prior works in exploring what a broader, comprehensive, and iterative approach to the assessment process addressing the \textit{normative core} could look like.

Principles alone cannot guarantee that the impacts and unintended consequences of A/IS are being identified and adequately addressed \cite{mittelstadt2019principles}. Oftentimes, nominal adoption of ethical principles can even dilute responsibility and lead to the so-called \textit{ethics washing} within A/IS organisations \cite{bietti2020ethics}. Various scholars have suggested the use of well-being metrics in the evaluation of A/IS impacts \cite{musikanski_ieee_2018, paradigm}. The focus on well-being in this area of research has been motivated by (1) the development of our ability to measure well-being \cite{diener1999subjective, oecd2013oecd, durand2015oecd, musikanski2017happiness}, in part as an alternative to a traditional focus on the paradigm of economic growth \cite{ieeeWhatMatters, kubiszewski2013beyond, reiser2011benefit}, as well as (2) the increased urgency of resolving pressing issues of misuse and unintended negative consequences of A/IS \cite{o2016weapons, zuboff2019age, russell2019human}. Schiff, Murahwi, Musikanski, and Havens \cite{paradigm} propose that A/IS assessments need to incorporate (1) a broad-based definition of well-being, (2) user and stakeholder engagement, (3) rigorous but flexible indicators, (4) repeated data collection, (5) learning from well-being data, and (6) an iterative process. Hereby, we go on to explore how an A/IS EWIA process can functionally address key aspects of AI principles such as: privacy, accountability, safety and security, transparency and explainability, fairness and non-discrimination, human control of technology, professional responsibility, and the promotion of human values, i.e. the  "normative core" identified by Fjeld et al. \cite{fjeld2020principled}. Futhermore, we demonstrate how a broader, comprehensive, and iterative approach to the assessment process may be put in practice through examples from real-world A/IS use-cases.

Acknowledging the limitations of metrics-based approaches \cite{10.2307/j.ctvc77h85}, we do not propose that the EWIA is to replace all other approaches to ensure that A/IS is ethical, trustworthy, and beneficial. On the contrary, our approach is proposed as complementary, in particular to those exploring "ethics by design" discussed by Virginia Dignum \cite{dignum2018ethics} or the "assessment list" for AI developers and deployers proposed by the European Commission High-Level Expert Group on AI \cite{assessmentList}. 

\input{01_section_wia_process}
\input{02_section_ai_principles_wia_indicators}

\input{03_multi_stakeholder}
\input{04_causality}
\input{05_conclusion}

\bibliographystyle{IEEEtran}
\bibliography{references}

\end{document}

%% file: 01_section_wia_process.tex
\section{A Well-being Impact Assessment Process}

We define well-being based on the Organization of Economic Cooperation and Development’s (OECD) well-being framework which includes “people’s living conditions and quality of life today (current well-being), as well as the resources that will help to sustain people’s well-being over time (natural, economic, human and social capital)” \cite{oecdMeasuring}. 
Well-being, defined subjectively, includes such components as flourishing, positive and negative affect, and satisfaction with life. A well-being paradigm shift in the evaluation of A/IS has been put forth by the IEEE P7010 Recommended Practice for Assessing the Impact of Autonomous and Intelligent Systems on Human Well-Being (P7010) \cite{ieeep7010}. P7010 proposes "well-being metrics relating to human factors directly affected by intelligent and autonomous systems and establishes a baseline for the types of objective and subjective data these systems should analyze and include (in their programming and functioning) to proactively increase human well-being" \cite{ieeePAR}.  In what follows, we demonstrate how the EWIA could be practically utilized to assess alignment with AI guidelines and principles.

Next, we draw from the regulatory impact assessments which \cite{/content/publication/7a9638cb-en} help policymakers assess the impact and consequences of planned policy action. The impacts are usually considered in terms of benefits and costs at the societal level for a given country. It also includes the distribution of the costs and benefits to different stakeholders. Apart from socio-economic analysis it explicitly includes analysis of environmental impacts. The involvement of stakeholders in the definition of impacts should be an inherent part of the process \cite{oecd2008introductory}.

In line with these impact assessment approaches, which constitute the core of our methodology, the EWIA is based on the identification of stakeholders, related impacts, and relevant metrics for the identified impacts. A comprehensive dashboard can be created based on the metrics. The main aspects of the process are:

\begin{itemize}
    \item Stakeholder-engagement: A wide range of stakeholders need to be engaged in order to identify potential impacts. The engagement should not be limited to users only. It is often crucial to involve experts from academia and non-governmental actors, e.g. human rights groups. The engagement should not be limited to one-off identification of (potential) impacts. Stakeholders need to be engaged in an iterative manner jointly revisiting the metrics on a regular basis.
    
    \item Iterative: The list of impacts is revisited regularly in order to check for the fit of the metrics with expected impacts and for identification of additional metrics for previously unidentified (and unintended) impacts. The EWIA incorporates continuous feedback loops between accountable stakeholders and intermediary steps in the A/IS life-cycle. 
    
    \item Life-cycle approach: The EWIA captures the whole life-cycle of the A/IS. It combines ex-ante and ex-post assessments informing continuous improvement of the A/IS towards positive well-being outcomes. 
\end{itemize}

%% file: 02_section_ai_principles_wia_indicators.tex
\section{Addressing Key Aspects of AI Principles through Well-being Impact Assessments}

We now explore how the proposed process and framework could be utilized to address the major themes in the emerging consensus on fundamental ethical principles for A/IS as discussed by Fjeld et al \cite{fjeld2020principled}. Beyond discussing the framework in general terms, we give preliminary 
examples of metrics relevant to each of the AI principles. The metrics and indices outlined below are to be understood as examples for illustration only. 

To investigate the utility of the EWIA, we will contextualize our discussion through an example scenario. Company X Inc. is a large multinational organization whose leadership has been actively involved in spreading awareness of the need to incorporate ethics and rights-based perspectives in the whole A/IS system lifecycle. Currently, the company is developing a personal AI assistant to help parents educate and entertain their children. Using AI algorithms, the assistant proposes common activities for parents and children such as games and texts to be read to the children as well as online content for the consumption by children. The company recognizes the opportunity to significantly help parents and children, particularly in disadvantaged families, resulting in step change, braking path dependency, and allowing children to reach their full potential. However, the company also considered important risks such as reducing the time parents spend with their children or creating digital addition. They recognize the need "to develop [a] better understanding of not just the impact of screen time as a whole, but also between different types of screen time and children's development and wellbeing" \cite{onlineharms}.  Therefore, the company has adopted the \textit{normative core} of AI principles discussed by Fjeld et al. \cite{fjeld2020principled} at the organisational level. To put the principles into practice, they are adopting the EWIA framework through ex-ante and ex-post assessments with the aim to fully integrate them into the whole A/IS lifecycle. The stakeholders that they work with include regulators, local government, local nurseries, kindergartens, and civil society groups which operate in the city where the technology is going to be piloted. Parents are also included as a stakeholder in the EWIA and invited to engage in the assessments through participatory methods \cite{Slavin2016Design}. 

\subsection{Privacy}
Privacy is directly related to fundamental human rights. As discussed by various scholars, privacy-related AI principles and recommendations stipulate that the privacy of individuals is respected and that individuals maintain autonomy over the decisions regarding the use of their data \cite{fjeld2020principled, chander2019catalyzing}. Many A/IS cannot function in a beneficial manner without exploring often very detailed data about individuals \cite{solove2008understanding}. Data collection is an inherent part of the EWIA as proposed by IEEE's well-being assessment process \cite{ieeep7010, schiff2020ieee}. Privacy becomes of urgent importance and needs to be addressed through data and algorithmic governance models that could be integrated with the EWIA. To include privacy into the EWIA, multiple metrics could be defined to measure users' awareness of which personal data are collected and how are they used by the A/IS. Another index or metric could monitor technical solutions to privacy, such as secure multiparty computation \cite{russell2019human} or privacy-preserving deep learning \cite{shokri2015privacy}. Privacy and data governance is one of the requirements identified in the guidelines of the European Commission’s High-Level Expert Group on Artificial Intelligence \cite{ecHLAI}. Some A/IS are subject to legislation such as the European Union's General Data Protection Regulation, which defines explicit rules about data and privacy protection. A concrete example of the EWIA regarding privacy may be related to the monitoring and assessment that only data directly needed to fulfil the system’s objective are collected and stored. As stressed several times in this paper, the EWIA is proposed to supplement other approaches, not replace them. In the case of privacy-related AI principles, EWIAs could enrich the concept of privacy-by-design by facilitating the continuous monitoring of A/IS privacy considerations after A/IS are deployed. 

\subsection{Accountability}
Algorithmic accountability relates to our ability to assign responsibility for harm when algorithmic decision-making results in malicious outcomes or leads to unintended consequences such as discrimination and inequitable outputs. Drawing from the field of Science and Technology Studies (STS) we aim to also bring forward the concept of \textit{distributed agency} \cite{bruni2007reassembling} which is closely related to responsibility and accountability.  Recent efforts, such as the Annotation and Benchmarking on Understanding and Transparency of Machine Learning Lifecycles \cite{raji2019ml} project, could aid in closing the accountability gaps through the adoption of documentation practices. The documentation practices could be adopted and monitored as part of the EWIA by including them as key-performance indicators and performance evaluation metrics for the individuals and teams tasked with the design and development of A/IS. Furthermore, we pose that these documentation practices should encompass the A/IS impact assessment process through a dynamic and iterative approach. In addition, the EWIA framework could facilitate the system’s auditability by internal and/or independent actors, by ensuring traceability and logging of the AI system’s processes and outcomes.

\subsection{Safety and security}
Concerns about the safety and security of AI systems have been the main area of research in the nascent field of AI Safety. Still, the meaning of safety is not well defined in different contexts. The European Commission’s High-Level Expert Group on Artificial Intelligence discusses the safety of a system in close relation to its reliability - “the system will do what it is supposed to do without harming living beings or [its] environment” \cite{ecHLAI}. Similarly, safety is often discussed as the problem of accidents in AI systems “defined as unintended and harmful behavior that may emerge from poor design of real-world AI systems” \cite{amodei2016concrete}. Other scholars have discussed safety in terms of being able to address the challenges of specification, robustness, and assurance of AI systems \cite{leike2017ai}. Brundage et al. \cite{brundage2018malicious} have proposed a taxonomy of AI security domains including digital, physical, and political security. 

Building on various approaches in measuring the quality of cybersecurity \cite{globalCybersecurityIndex}, a safety-security index could be developed. In particular, the index could include technical measures and organizational measures. These could also cover effective fallback plans along similar lines as proposed by the European Commission’s High-Level Expert Group on Artificial Intelligence \cite{ecHLAI}.

The common ground among these prior works is that the safety and security of a system depend on the sociotechnical context within which the system exists \cite{selbst2019fairness}. Therefore, there’s a need for a critical approach to evaluating safety in a broader context which could be enabled by an ecosystem that supports external impact assessments \cite{solaiman2019release, rakova2020assessing}. In the context of existing policy frameworks, safety and security are closely related to community, government, and human rights. An additional metric inspired by well-being indicators could be the measurement of community safety, defined as going about “daily life without fear or risk of harm or injury” \cite{safeNZ} could help A/IS practitioners to monitor user-perceived safety. For example, an autonomous vehicles A/IS could employ such a metric as part of their EWIA in order to measure the perceived safety of the technology.

\subsection{Transparency and explainability}
An A/IS is often referred to as a black box system and perceived in terms of its inputs and outputs, without any knowledge of its internal workings. Safeguarding of intellectual property, competitive markets, and other incentives might disincentivize A/IS creators to "open" a black box by providing sufficient transparency about the operations of their system. IEEE’s Ethically Aligned Design recommends the development of "new standards that describe measurable, testable levels of transparency, so that systems can be objectively assessed and levels of compliance determined" \cite{ieee2016ethically}. However, various scholars have argued that transparency is not a required property for A/IS \cite{rudin2019stop, wachter2017counterfactual}, instead, there’s a growing need for a sufficient level of explainability and interpretability \cite{sokol2020explainability}. Regulatory frameworks such as the EU’s General Data Protection Regulation (GDPR) provision that a right to explanation is a key step towards algorithmic accountability \cite{kaminski2019right}. We posit that the EWIA could help in evaluating the technical level of transparency as well as the efficacy of the interface between people and A/IS, specifically assessing if a sufficient level of explainability and interpretability is achieved. 
Although transparency and explainability of A/IS bring about complex technical challenges, an EWIA metric could capture transparency about the A/IS objective(s), for example, measuring users' awareness of the perceived A/IS objective(s) \cite{hoffman2018metrics}.

\subsection{Fairness and non-discrimination}
Scholars have argued that narrow technical conceptualizations of fairness and discrimination have far-reaching implications, instead, there’s a growing need for a process-driven approach to the evaluation of fairness and non-discrimination \cite{barabas2020studying}. An iterative EWIA process could ensure that appropriate fairness criteria are integrated into the ML life-cycle instead of being reduced to a mathematical property of algorithms, independent from specific social contexts. In addition, a subjective perception of users about fairness and non-discrimination could be taken into account. A related  metric could be inspired by a well-being indicator such as the sense of discrimination in one’s neighborhood or community \cite{b111}. 


We pose that when adequate EWIA indicators are employed throughout the A/IS life-cycle, A/IS creators will be able to discover and mitigate potential fairness and discrimination issues ex-ante in the development stage as well as during the continuous monitoring after the system is deployed. Technical metrics could measure the quality of training data in terms of its representativeness regarding the given population where the A/IS is being deployed \cite{gebru2018datasheets,holland2018dataset}. 

\subsection{Human control of technology} 
The principles under this theme require that important decisions remain subject to human review \cite{fjeld2020principled}. Human control of technology is also among the main principles outlined by social scientists and philosophers: “A system that understands us better than we understand ourselves can predict our feelings and decisions, can manipulate our feelings and decisions, and can ultimately make decisions for us” \cite{harari}. There is a need to define metrics for various situations which would help examine the behavioral impacts of A/IS. Moreover, a metric has to monitor that the system is not manipulating preferences or creating addictive behavior \cite{russell2019human, rakova2019human}.
Other metrics could monitor how users perceive the ease of opting out of an automated decision \cite{fjeld2020principled}. Nudging by A/IS should be based on explicit consent and opt-in mechanisms \cite{ieee2016ethically}. Similarly, related metrics could monitor that users are aware when the system is trying to directly influence their behavior as well as measuring how easily users can access the human review of automated decisions.


\subsection{Promotion of human values}
As discussed by Fjeld et al., this general principle includes human values and human flourishing, leveraged to benefit society, and access to technology and benefits thereof \cite{fjeld2020principled}. These principles can be directly covered by the EWIA as they are part of traditional well-being measurement frameworks \cite{musikanski2017happiness}. In particular, metrics related to satisfaction with life, affect, psychological well-being, and inequality may be utilised to assess the implementation of this general principle in a concrete A/IS deployment. Futhermore, metrics could be could be rooted in existing frameworks for human rights assessments \cite{abrahams2010guide}, enabling the consideration of the human rights of affected stakeholders (people, groups, or communities) at each stage of the A/IS lifecycle.



\subsection{Professional responsibility} 
Professional responsibility A/IS principles evolve around considerations of long-term effects and responsible design. They are grounded in recognizing "the vital role that individuals involved in the development and deployment of AI systems play in the systems’ impacts, and call on their professionalism and integrity in ensuring that the appropriate stakeholders are consulted and long-term effects are planned for" \cite{fjeld2020principled}.
It seems that the overall concept of EWIA directly responds to the aspects raised under this principle. The EWIA is meant to help envisage, monitor, and assess the long term effects in a multi-stakeholder manner. It is also focused on enabling the professionals who develop and deploy A/IS to better understand the impacts these systems may have. Some of the previously proposed approaches based on checklists \cite{madaio2020co} may not sufficiently cover the continuum between binary decision and related trade-offs which may create significant application challenges for practitioners. The checklist approach also does not envisage continuous monitoring over time, in particular, in the deployment phase of the A/IS life-cycle. Although we do not suggest that the EWIA replaces all other approaches, systematic measurement of a comprehensive set of metrics could significantly improve the positive effect of existing and new AI governance mechanisms.

%% file: 03_multi_stakeholder.tex
\section{Multi-stakeholder monitoring and assessment systems}

In the previous sections, we have discussed the potential of the EWIA to positively contribute to the practical implementation of fundamental AI principles. The intermediate product of EWIAs would be a set of metrics to be monitored at variable frequencies throughout the A/IS life-cycle. In order to enable such monitoring and assessment of the results, a dedicated system needs to be developed. 

There’s an emergent need for new kinds of testing infrastructure in the form of monitoring and assessment systems supported by enabling organizational structures that could facilitate the EWIA implementation. Creating such testing infrastructure may be beyond the abilities of traditional A/IS organizations due to (1) the high complexity of A/IS impacts and their measurement, (2) the need for comprehensive approach to impacts identification and assessment, (3) the need for in-depth expertise in many areas, (4) representation of diverse perspectives, and also (5) related costs for developing and maintaining these kinds of monitoring and assessment systems. Governments or non-profit actors may be better placed to co-develop such testing infrastructures and offer them for free and voluntary use by companies and other actors considering the deployment of A/IS. Third party actors can also help to ensure the quality control of such systems.

The monitoring and assessment system would assist the A/IS entity to specify which EWIA data need to be gathered, how often it is going to be collected (as there will be different time spans for different metrics), and how it is going to be analyzed in order to understand related impacts, both intended and unintended. A multistakeholder approach would also ensure a higher degree of comprehensiveness when compared to approaches deployed by individual companies. Based on the discussion in this paper, we posit that comprehensiveness is one of the most important requirements for an effective EWIA. Without a comprehensive approach to A/IS assessments, crucial first and second order impacts may be neglected. In turn, it would be near impossible  for decision-makers to identify potential important trade-offs among the impacts.

Security of data and safeguarding of intellectual property rights are among the main prerequisites for such a system. The results of monitoring and assessment should at first be available to companies with confidential reporting mechanisms. It is important to acknowledge that there may be serious uncertainties about identified impacts. It is imperative to build the systems on the basis of high trust. Institutions with high credibility and trust such as the OECD, IEEE, or consumer organisations may be among the entities to support such systems.

In the future, it would be in the interest of public actors to gain a much better understanding of potential impacts, and to create a testing environment of potential regulatory action. Should regulatory action be deemed necessary, the EWIA coulds allow for regulation to be grounded in comprehensive evidence, in line with evidence-based policy making and regulatory impact assessment processes. 

The main motivation for the private sector to voluntarily take advantage of such monitoring systems would be related to reputation and general risk management. The multi-stakeholder system would allow for assessing the impact on well-being of A/IS pilot implementations. Developers of A/IS would be able to gather important insights about the impacts of artefacts they create. By assessing deployment of novel A/IS via the EWIA, it would be possible to identify potential harm early on and consider and implement mitigation strategies. Such an approach would also allow for the measurement of success vis-a-vis mitigation measures. Finally, some proposed regulatory requirements \cite{wright2019online} will ask companies to improve their understanding of the risks associated with their service. The EWIA could play an important role in fulfilling such a duty.

%% file: 04_causality.tex
\section{Understanding of causality through impact assessments} 
A well-being assessment-based approach could allow for measuring benefits as well as harms to users and other stakeholders (also including future generations via its focus on sustainability). Although it is often possible to assign contribution of effects of A/IS, oftentimes, the causal links between the operation of A/IS and the observed effects are much more difficult to develop. Recognizing the immense complexity of the socio-technical context within which A/IS are situated, we think that this should not hinder further investigations. Often, discussed negative impacts, such as erosion of democracy caused by \textit{echo-chambers} powered by algorithms
, may serve as an example where the causal links are not clear. However, it is crucial to understand causal links related to negative impacts not only at the level of actual users but also in the wider context of impacts at the societal level.

In line with regulatory impact assessments and an evidence-based approach to policy-making, we need to improve the understanding of causal links between observed impacts and a given A/IS. If evidence is not available, decision-makers should use available methods to find out what works. 
Similarly, we need to tackle the attribution challenge establishing cause and effect between A/IS and observed effects. To tackle this challenge, the assessment should be capable of addressing inter alia confounding factors, spillover effects, and impact heterogeneity by intervention, beneficiary, and context \cite{ien}.

Policy impact evaluation designs use a control group to allow for confounding factors \cite{ien2}. That is, apart from the measurement of impacts, we need a counterfactual: what would have occurred if the A/IS had not been implemented? A traditional approach based on Randomized Controlled Trials (RCTs) may be an inspiration for improving the understanding between the A/IS and its impacts \cite{martinez2011impact}. Claims about cause and effect can be made with more confidence when they are based on findings of randomized trials, rather than on almost any other type of study \cite{martinez2011impact}. Proposed multi-stakeholder EWIA monitoring and  assessment system could also allow the studying of pilot use-cases of new or updated A/IS. These designs may include well-designed A/B testing where, for example, one group of users interacts with a new recommender system (or new features of an existing system) while another does not. Assessing the data not only where A/IS was deployed but also from relevant reference non-implementation sites (control group), in line with the RCT as well as quasi-experimental methodologies, designers could improve the assignment of causality between the AI system and its observed impacts. The unit of analysis may be an individual user, a group of users, geographical areas (e.g. the A/IS is implemented only in one of two similar regions), etc.

Future regulatory approaches are expected to be grounded in evidence. 
Due to the immense speed of technology development and its widespread deployment, such regulatory actions will need to be dynamic and comprehensive. The iterative and multi-stakeholder nature of EWIAs could allow for and enable AI governance that is rooted in collaboration and cooperation.

%% file: 05_conclusion.tex
\section{Conclusion and Future Work}

We have illustrated how well-being impact assessment of A/IS could be enhanced to capture, at least partially, the \textit{normative core}, i.e. emerging consensus on ethical and rights-based AI principles. Next, we have outlined a multi-stakeholder monitoring and assessment system which would allow for putting the enhanced set of metrics into practice. Establishing and launching the monitoring and assessment system would also enable validation of the proposed approach. In particular, by providing sets of values regarding the impacts which could be in turn verified by appropriate social sciences methods. We  do not propose that the EWIA approach should replace other methods and approaches such as various organizational measures, ethics codes, certification schemes, or standards, but instead to complement them. EWIAs enabled by multi-stakeholder monitoring and assessment systems could also help highlight longer-term impacts of A/IS in particular to developers and corporate boards. This seems important as we humans are prone to myopic behavior and often do not assign sufficient value towards future impacts of our decisions.

In addition, future work could consider how the EWIA might serve as a basis for a complex automated approach to the assessment of A/IS impacts, while maintaining full explainability of the findings. For example, more attention will have to be paid to the appropriation of impacts to an individual A/IS. Here we have suggested that an approach inspired by impact assessments in policymaking could be the way forward. Furthermore, we will need to focus research on navigating difficult trade-offs, e.g. between safety and privacy or human control and accessibility. Lastly, apart from the assessment at the level of individuals and groups we also need to better understand what impacts AI deployment may have at a societal level, in particular regarding social cohesion and democracy. We hope that the enhanced well-being impact assessment discussed here could help in closing the gaps between principles and practice by elucidating a well-being-centered approach to potential negative effects and unintended consequences of A/IS through transparent and interpretable means.